\documentclass[aps,prl,twocolumn,showpacs]{revtex4-1} 

\usepackage{amsmath,amssymb,mathrsfs}
\usepackage{latexsym}
\usepackage{graphicx} 
\usepackage{epstopdf}
\usepackage{graphicx,epstopdf,color}
\usepackage{amsfonts}
\usepackage{bm}

\newcommand{\bs}[1]{\boldsymbol{#1}}
\newcommand{\be}{\begin{equation}}
\newcommand{\ee}{\end{equation}}
\newcommand{\bea}{\begin{eqnarray}}
\newcommand{\eea}{\end{eqnarray}}

\renewcommand{\phi}{\varphi}
\renewcommand{\epsilon}{\varepsilon}

\begin{document}

\title{Renormalization group analysis of competing quantum phases in the $J_1$-$J_2$ Heisenberg model on the kagome lattice}

\author{Raik Suttner${}^1$}
\author{Christian Platt${}^1$}
\author{Johannes Reuther${}^2$}
\author{Ronny Thomale${}^3$}
\affiliation{${}^1$Institute for Theoretical Physics and Astrophysics, University of W\"urzburg, D 97074 W\"urzburg}
\affiliation{${}^2$Department of Physics, California Institute of Technology, Pasadena, California 91125, USA}
\affiliation{${}^3$ Institut de th\'eorie des ph\'enom\`enes physiques, \'Ecole Polytechnique F\'ed\'erale de Lausanne (EPFL), CH-1015 Lausanne}

 \pagestyle{plain}

\begin{abstract}
Recent discoveries in neutron scattering experiments for Kapellasite and Herbertsmithite as well as theoretical calculations of possible spin liquid phases have revived interest in magnetic phenomena on the kagome lattice.
We study the quantum phase diagram of the $S=1/2$ Heisenberg kagome model as a function of nearest neighbor coupling $J_1$ and second neighbor coupling $J_2$. Employing the pseudofermion functional renormalization group, we find four types of magnetic quantum order (${\bf q}=0$ order, cuboc order, ferromagnetic order, and $\sqrt{3}\times\sqrt{3}$ order) as well as extended magnetically disordered regions by which we specify the possible parameter regime for Kapellasite. 
In the disordered regime $\frac{J_2}{J_1}\ll 1$, the flatness of the magnetic susceptibility at the zone boundary which is observed for Herbertsmithite can be reconciled with the presence of small $J_2>0$ coupling. In particular, we analyze the dimer susceptibilities related to different valence bond crystal (VBC) patterns, which are strongly inhomogeneous indicating the rejection of VBC order in the RG flow.  
\end{abstract}


\maketitle

{\it Introduction.} Frustrated magnetism is a focus of contemporary research in condensed matter physics, combining a plethora of experimental scenarios and diverse theoretical approaches to describe them. One of the most fascinating challenges of the field has been to investigate and understand the interplay of magnetic quantum order and disorder on the kagome lattice. A major reason why this lattice of corner-sharing triangles yields such a complicated structure of quantum phases is already evident from the classical kagome Heisenberg model (KHM): As a function of nearest neighbor and next nearest neighbor Heisenberg couplings $J_1$ and $J_2$, many different magnetic orders are present~\cite{harris-92prb2899}, where an infinite number of degenerate ground states can be found~\cite{chalker-92prl855}. From a theoretical perspective, not many rigorous results about the quantum phase diagram are known so far. Advanced mean field theories have provided important guidance as to what type of ordered and disordered quantum phases could possibly be found~\cite{anderson87s1196,marston-91jap5962, wen91prb2664,sachdev92prb12377,hastings00prb014413,ran-07prl117205}, but cannot give unambiguous information about which phase will eventually be stabilized in the microscopic model. A peculiar feature of the KHM which is known since early exact numerical calculations of finite size clusters~\cite{waldtmann98epj501} is the large amount of singlet states at low energy. This suggests a plethora of  competing quantum-disordered phases and is probably one of the main reasons why the interpretation of present results for the $J_1$ KHM from microscopic numerical approaches is not yet settled~\cite{jiang-08prl117203,yan-11s1173,depenbrock-12prl067201,iqbal-cm1209}. 

Transferring our fragile theoretical knowledge to experimental scenarios is even more challenging. The Herbertsmithite compound  ${\mathrm{ZnCu}}_{3}(\mathrm{OH}{)}_{6}{\mathrm{Cl}}_{2}$ is one of the rare properly investigated material realizations of a $S=1/2$ kagome spin model, which is supposedly dominated by the $J_1$ term.
Early investigations from neutrons~\cite{helton-07prl107204} and muon spin rotation~\cite{mendels-07prl077204} have already indicated its unconventional frustrated magnetic properties, exhibiting no sign of magnetic order down to a few mK.
While there is no indication for a magnetic spin gap, it is likely that this does not hint at a key property of a possible KHM description, but might be due to Dzyaloshinskii-Moriya (DM) effects and impurities which further complicate the picture~\cite{elhajal-02prb014422,helton-07prl107204,devries-08prl157205,helton-10prl147201,han-12prl157202}. Detailed latest neutron scattering experiments resolve an extremely flat magnetic susceptibility profile~\cite{wulferding-10prb144412,han-12n406}, which allows for an interpretation along spin fractionalization as known from spinon continua in the spin structure factor of quasi-1d systems~\cite{coldea-03prb134424}. Recently, neutron and $\mu$SR experiments on Kapellasite ${\mathrm{Cu}}_{3}\mathrm{Zn}(\mathrm{OH}{)}_{6}{\mathrm{Cl}}_{2}$, of which Herbertsmithite is a polymorph, have found indication for cuboc magnetic order~\cite{flaak-12prl037208}. This is an exciting discovery, as electronic structure calculations find dominant {\it antiferromagnetic} $J_1$ coupling and hence a similar regime as Herbertsmithite~\cite{janson-08prl106403}. From the knowledge where classical cuboc order emerges, however, this suggests the presence of {\it ferromagnetic} $J_1$ and antiferromagnetic $J_2$ in Kapellasite. 

\begin{figure*}[t]
\centering
$\begin{matrix}
\begin{matrix}\includegraphics[scale=0.16]{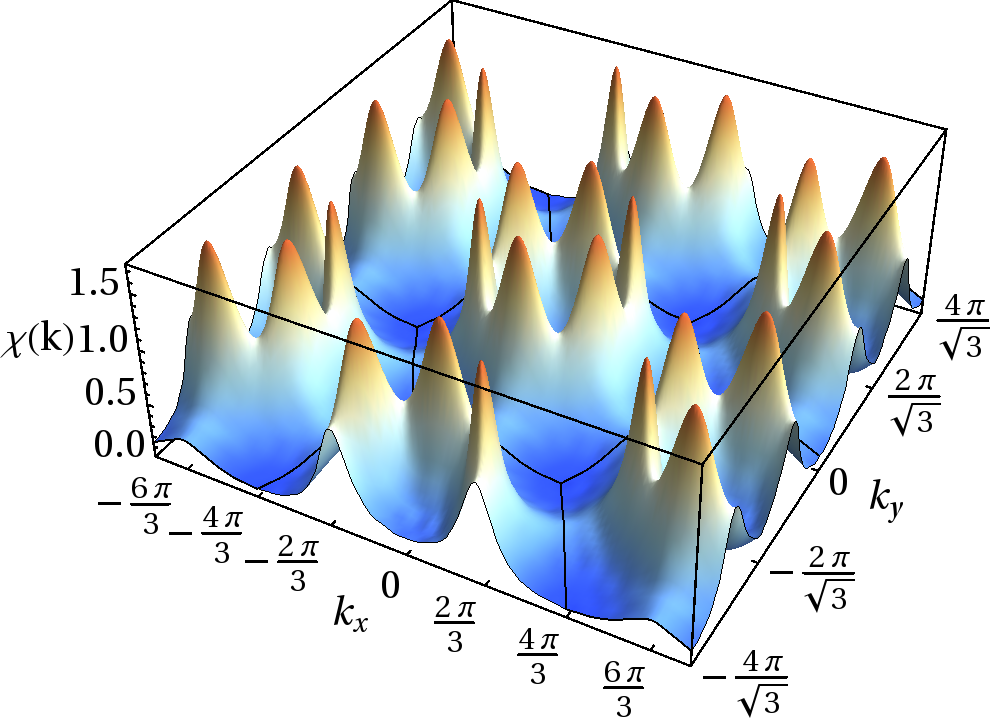}\end{matrix}\!\!\!\!\!\!\!\!\!\!\!\!\!\!\!\!\!\!\!\!\!\!\!\!\!\!\!\!\!\!\!\!\!\!\!\!\!\!\!\!\!\!\!\!\!\!\!\!\!\!\!\!\!\!\!\!\!\!\!\!\!\!\!\!\!\!\!\!\!\!\!\!\!\!\!\!\!\!\!\!\!\!\!\!\!\!\!\!\!\!\!\!\!\begin{matrix}\text{cuboc}\\\vphantom{m}\\\vphantom{m}\\\vphantom{m}\\\vphantom{m}\\\vphantom{m}\\\vphantom{m}\\\vphantom{m}\\\vphantom{m}\\\vphantom{m}\end{matrix}\qquad\qquad\qquad\qquad\qquad\qquad & \quad\begin{matrix}\includegraphics[scale=0.8]{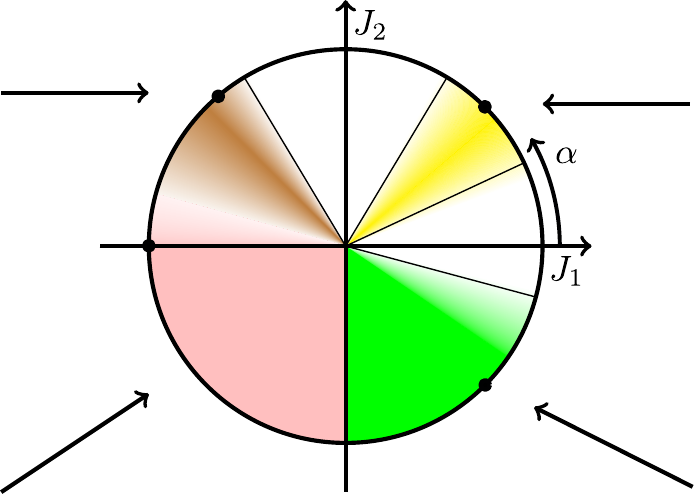}\end{matrix} & \!\!\!\!\!\begin{matrix}\includegraphics[scale=0.16]{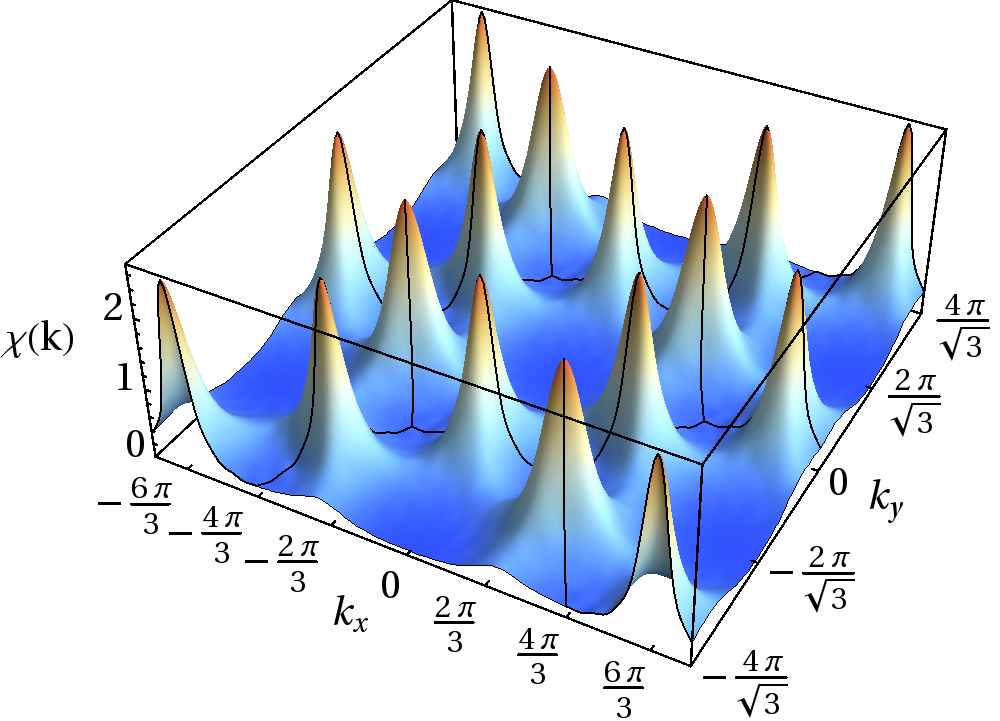}\end{matrix}\!\!\!\!\!\!\!\!\!\!\!\!\!\!\!\!\!\!\!\!\!\!\!\!\!\!\!\!\!\!\!\!\!\!\!\!\!\!\!\!\!\!\!\!\!\!\!\!\!\!\!\!\!\!\!\!\!\!\!\!\!\!\!\!\!\!\!\!\!\!\!\!\!\!\!\!\!\!\!\!\!\!\!\!\!\!\!\!\!\!\!\!\!\begin{matrix}\text{$\mathbf{q}\!=\!0$}\\\vphantom{m}\\\vphantom{m}\\\vphantom{m}\\\vphantom{m}\\\vphantom{m}\\\vphantom{m}\\\vphantom{m}\\\vphantom{m}\\\vphantom{m}\end{matrix}\qquad\qquad\qquad\qquad\qquad\qquad \\ \!\begin{matrix}\includegraphics[scale=0.16]{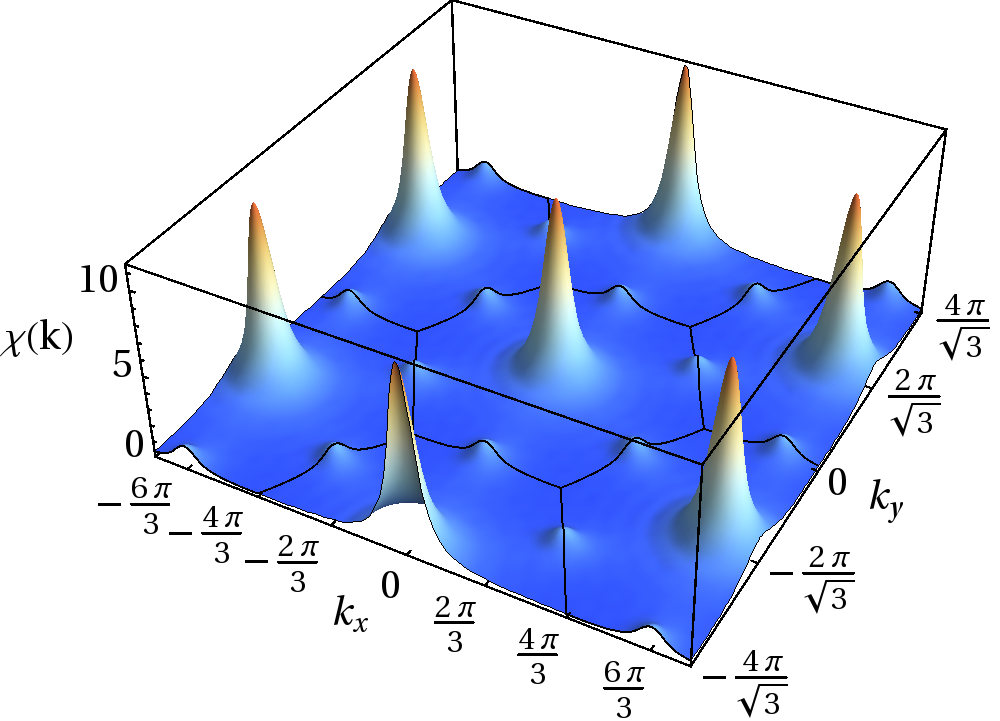}\end{matrix}\!\!\!\!\!\!\!\!\!\!\!\!\!\!\!\!\!\!\!\!\!\!\!\!\!\!\!\!\!\!\!\!\!\!\!\!\!\!\!\!\!\!\!\!\!\!\!\!\!\!\!\!\!\!\!\!\!\!\!\!\!\!\!\!\!\!\!\!\!\!\!\!\!\!\!\!\!\!\!\!\!\!\!\!\!\!\!\!\!\!\!\!\!\begin{matrix}\text{ferro}\\\vphantom{m}\\\vphantom{m}\\\vphantom{m}\\\vphantom{m}\\\vphantom{m}\\\vphantom{m}\\\vphantom{m}\\\vphantom{m}\\\vphantom{m}\end{matrix}\qquad\qquad\qquad\qquad\qquad\qquad & \quad\begin{matrix}\includegraphics[scale=0.15]{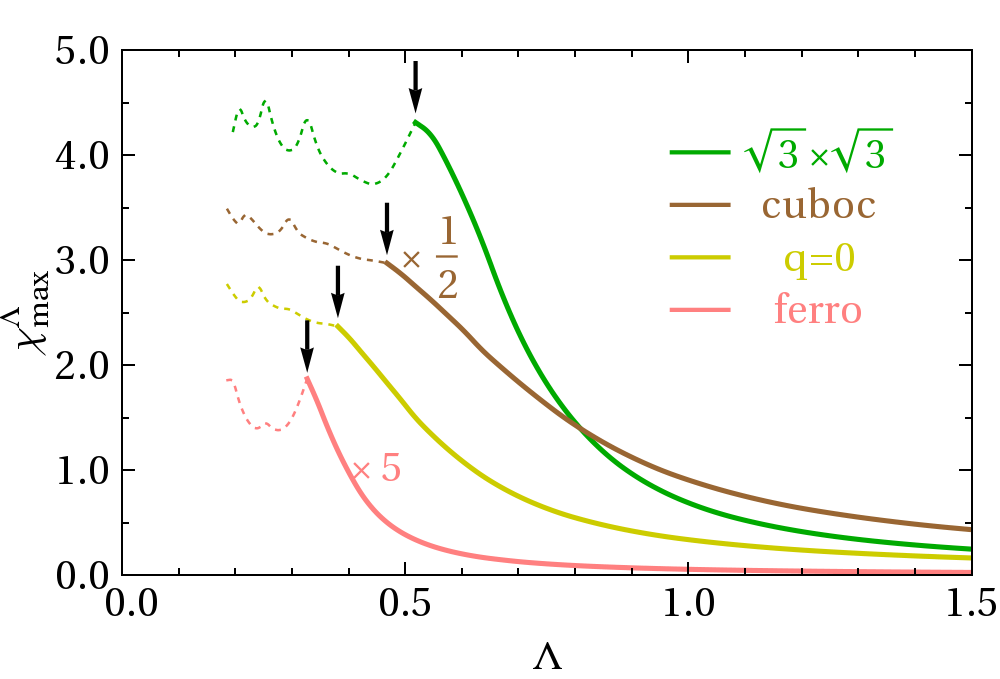}\end{matrix} & \begin{matrix}\includegraphics[scale=0.16]{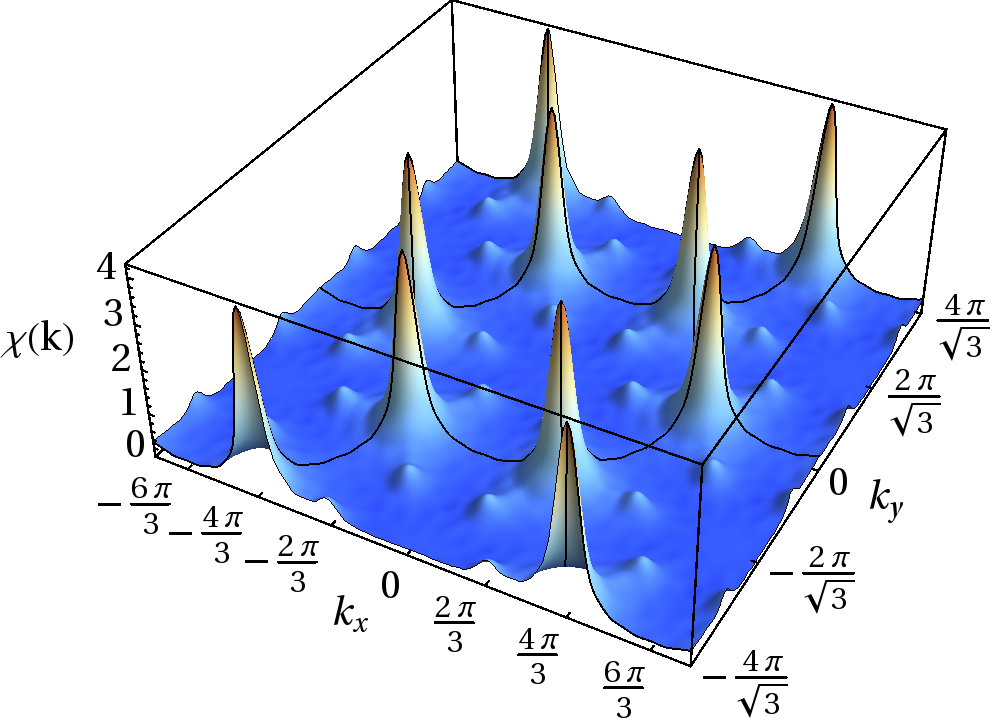}\end{matrix}\!\!\!\!\!\!\!\!\!\!\!\!\!\!\!\!\!\!\!\!\!\!\!\!\!\!\!\!\!\!\!\!\!\!\!\!\!\!\!\!\!\!\!\!\!\!\!\!\!\!\!\!\!\!\!\!\!\!\!\!\!\!\!\!\!\!\!\!\!\!\!\!\!\!\!\!\!\!\!\!\!\!\!\!\!\!\!\!\!\!\!\!\!\begin{matrix}\text{$\sqrt{3}\!\times\!\!\sqrt{3}$}\\\vphantom{m}\\\vphantom{m}\\\vphantom{m}\\\vphantom{m}\\\vphantom{m}\\\vphantom{m}\\\vphantom{m}\\\vphantom{m}\\\vphantom{m}\end{matrix}\qquad\qquad\qquad\qquad\qquad\qquad
\end{matrix}$
\caption{Top center: Phase diagram of the $J_1$-$J_2$ kagome model. Colored regions correspond to the magnetically ordered phases $\bs{q}=0$, cuboc, ferromagnetic and $\sqrt{3}\times\sqrt{3}$ order, white regimes are non-magnetic. Faint areas near the phase transitions (e.g. between the cuboc and the ferromagnetic phase) are regions of enhanced uncertainty. Bottom center: Flow of the susceptibilities in the four ordered regimes. Arrows indicate the instability features. The curves for $\bs{q}=0$, cuboc, ferromagnetic and $\sqrt{3}\times\sqrt{3}$ order are given for $\alpha=45^\circ$, $130^\circ$, $180^\circ$ and $315^\circ$, respectively (black dots in the phase diagram). The left and right sides of the figure show the susceptibility profiles for these types of orders (at the same values for $\alpha$) at the instability breakdown.}
\label{fig:magn_order}
\end{figure*}

\begin{figure*}[t]
\centering
$\begin{matrix}\begin{matrix} \includegraphics[scale=0.18]{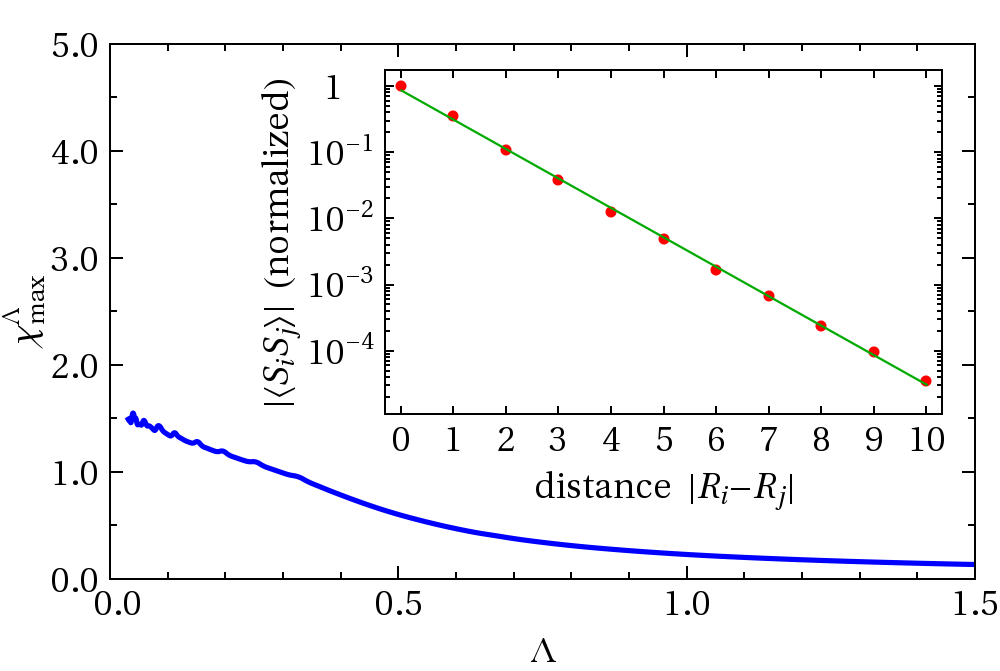} \end{matrix} \!\!\!\!\!\!\!\!\!\!\!\!\!\!\!\!\!\!\!\!\!\!\!\!\!\!\!\!\!\!\!\!\!\!\!\!\!\!\!\!\!\!\!\!\!\!\!\!\!\!\!\!\!\!\!\!\!\!\!\!\!\!\!\!\!\!\!\!\!\!\!\!\!\!\!\!\!\!\!\!\!\!\!\!\!\!\!\!\!\!\!\!\!\!\!\!\!\!\!\!\!\!\!\!\!\!\!\!\!\!\!\!\!\!\!\!\!\!\! \begin{matrix} \text{(a)} \\ \vphantom{m} \\ \vphantom{m} \\ \vphantom{m} \\ \vphantom{m} \\ \vphantom{m} \\ \vphantom{m} \\ \vphantom{m} \\ \vphantom{m} \end{matrix} \qquad\qquad\qquad\qquad\qquad\qquad\qquad\qquad\qquad\quad\ \begin{matrix} \includegraphics[scale=0.15]{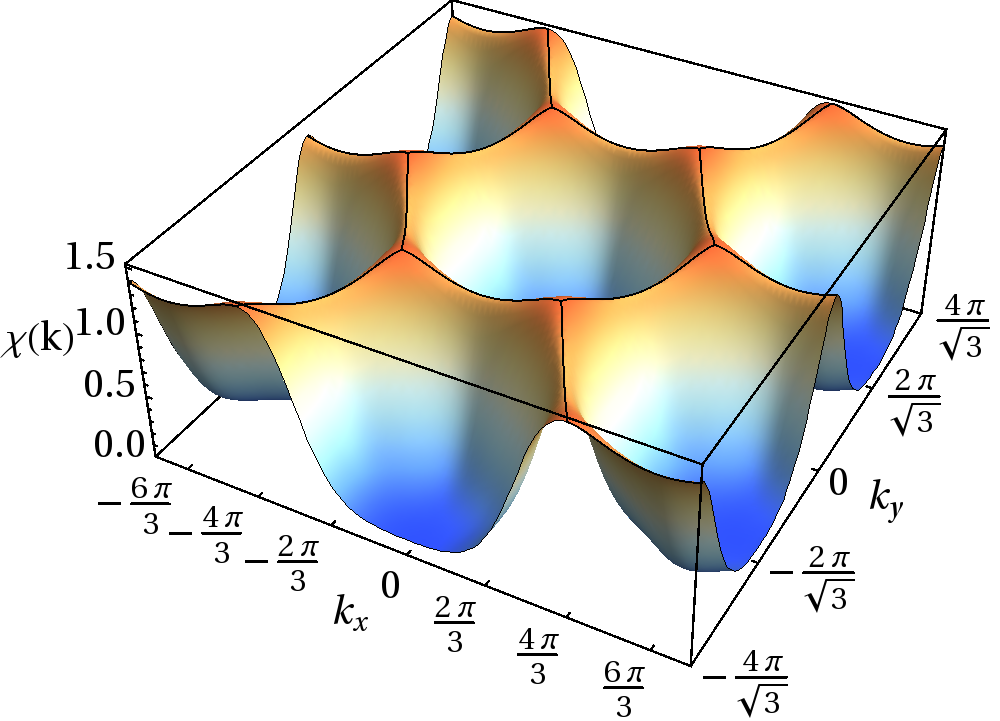} \end{matrix} \!\!\!\!\!\!\!\!\!\!\!\!\!\!\!\!\!\!\!\!\!\!\!\!\!\!\!\!\!\!\!\!\!\!\!\!\!\!\!\!\!\!\!\!\!\!\!\!\!\!\!\!\!\!\!\!\!\!\!\!\!\!\!\!\!\!\!\!\!\!\!\!\!\!\!\!\!\!\!\!\!\!\!\!\!\!\! \begin{matrix} \raisebox{0.3cm}{\text{(b)\hspace*{3cm}\raisebox{0.4cm}{$J_2=0$}}} \\ \vphantom{m} \\ \vphantom{m} \\ \vphantom{m} \\ \vphantom{m} \\ \vphantom{m} \\ \vphantom{m} \\ \vphantom{m} \\ \vphantom{m} \end{matrix} \hspace*{0.7cm} \begin{matrix} \includegraphics[scale=0.15]{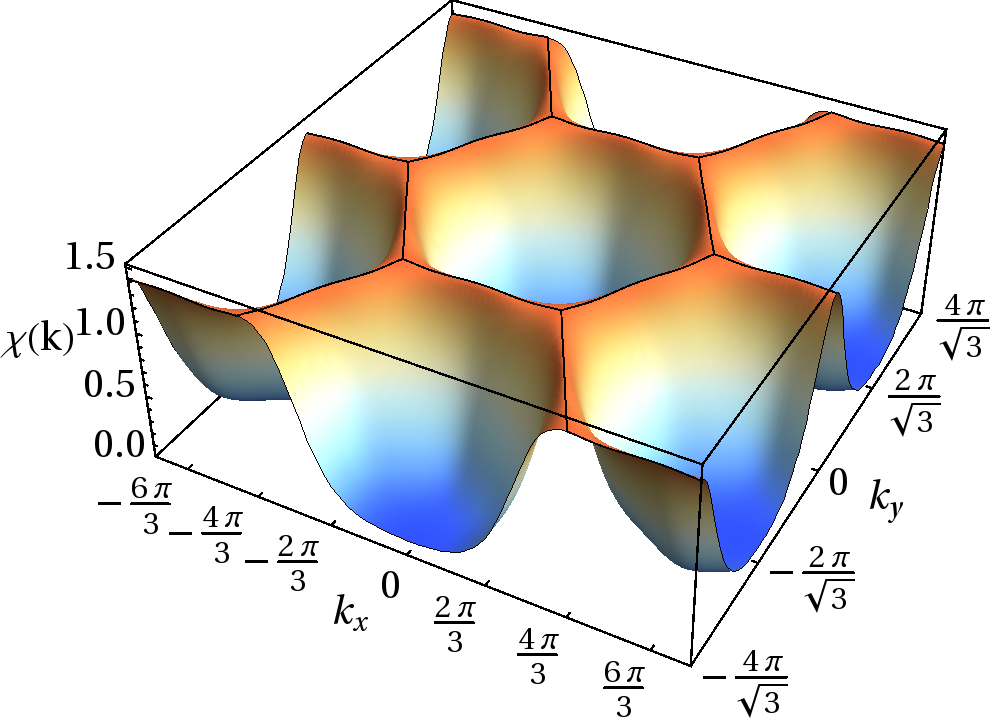} \end{matrix} \!\!\!\!\!\!\!\!\!\!\!\!\!\!\!\!\!\!\!\!\!\!\!\!\!\!\!\!\!\!\!\!\!\!\!\!\!\!\!\!\!\!\!\!\!\!\!\!\!\!\!\!\!\!\!\!\!\!\!\!\!\!\!\!\!\!\!\!\!\!\!\!\!\!\!\!\!\!\!\!\!\!\!\!\!\!\! \begin{matrix} \raisebox{0.3cm}{\text{(c)}\hspace*{2.8cm\raisebox{0.4cm}{$J_2=0.017J_1$}} } \\ \vphantom{m} \\ \vphantom{m} \\ \vphantom{m} \\ \vphantom{m} \\ \vphantom{m} \\ \vphantom{m} \\ \vphantom{m} \\ \vphantom{m} \end{matrix} \qquad\qquad\qquad\qquad\qquad\qquad\qquad\end{matrix}$
\caption{Non-magnetic phase for $\vert \frac{J_2}{J_1} \vert \ll 1$: (a) $\Lambda$-flow of the largest $\bs{k}$-component of the magnetic susceptibility for $\alpha=0$. The flow behavior is smooth to $\Lambda=0$. (Small oscillations below $\Lambda\approx0.3$ are due the frequency discretization.) (b), (c) $\bs{k}$-space resolved susceptibility at $\Lambda=0$. ($\chi$ and $\Lambda$ are given in units of  $J_1$.) Small maxima can be seen at the K points of the second Brillouin zone for $J_2=0$ in (b). These maxima vanish for at $J_2=0.017J_1$ in (c), with strong resemblance to~\cite{han-12n406}.}
\label{fig:kagome_nn}
\end{figure*}

\begin{figure*}[t]
\centering
$ \begin{matrix} \begin{matrix}\includegraphics{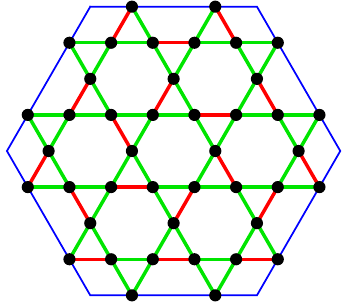}\end{matrix} \to \begin{matrix}\includegraphics{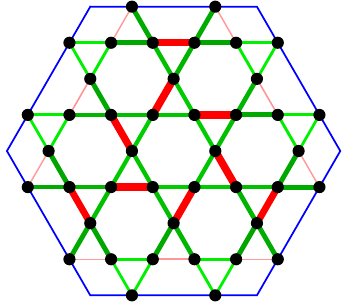}\end{matrix}\!\!\!\!\!\!\!\!\!\!\!\!\!\!\!\!\!\!\!\!\!\!\!\!\!\!\!\!\!\!\!\!\!\!\!\!\!\!\!\!\!\!\!\!\!\!\!\!\!\!\!\!\!\!\!\!\!\!\!\!\!\!\!\!\!\!\!\!\!\!\!\!\!\!\!\!\!\!\!\!\!\!\!\!\!\!\!\!\!\!\!\!\!\!\!\!\!\!\!\!\!\!\!\!\!\!\!\!\!\!\!\!\!\!\!\!\!\!\!\!\!\!\!\!\!\!\!\!\!\!\!\!\!\!\!\!\!\!\!\!\!\! \begin{matrix}\text{(a)}\\\vphantom{m}\\\vphantom{m}\\\vphantom{m}\\\vphantom{m}\\\vphantom{m}\\\vphantom{m}\end{matrix} \qquad\qquad\qquad\qquad\qquad\qquad\qquad\qquad\qquad\qquad\qquad\qquad\qquad\quad \begin{matrix}\includegraphics{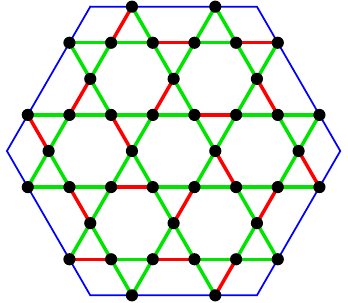}\end{matrix} \to \begin{matrix}\includegraphics{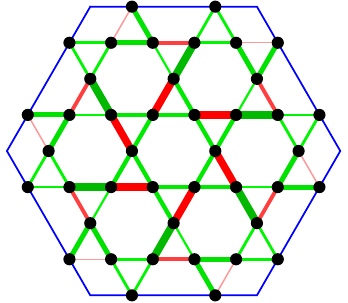}\end{matrix}\!\!\!\!\!\!\!\!\!\!\!\!\!\!\!\!\!\!\!\!\!\!\!\!\!\!\!\!\!\!\!\!\!\!\!\!\!\!\!\!\!\!\!\!\!\!\!\!\!\!\!\!\!\!\!\!\!\!\!\!\!\!\!\!\!\!\!\!\!\!\!\!\!\!\!\!\!\!\!\!\!\!\!\!\!\!\!\!\!\!\!\!\!\!\!\!\!\!\!\!\!\!\!\!\!\!\!\!\!\!\!\!\!\!\!\!\!\!\!\!\!\!\!\!\!\!\!\!\!\!\!\!\!\!\!\!\!\!\!\!\!\! \begin{matrix}\text{(b)}\\\vphantom{m}\\\vphantom{m}\\\vphantom{m}\\\vphantom{m}\\\vphantom{m}\\\vphantom{m}\end{matrix}\qquad\qquad\qquad\qquad\qquad\qquad\qquad\qquad\qquad\qquad\qquad \end{matrix}$
\caption{Dimer configuration for (a) the HVBC~\cite{singh-07prb180407} and (b) the modified HVBC before (left) and after (right) the RG flow. Red lines correspond to strengthened and green lines to weakened bonds, where the magnitude of the response is encoded in the thickness of the lines. The blue lines marks the unit cell of the pattern. (a) The local pinwheel structure gets enhanced during the flow, while the outer bond responses are inhomogeneous indicating the rejection of long-range order.
(b) A modified HVBC with $60^\circ$ rotation symmetry shows similar features as (a).}
\label{fig:dimer}
\end{figure*}

In this Letter, we develop a pseudofermion functional renormalization group (PFFRG) perspective on the {$J_1$-$J_2$} kagome Heisenberg model which is ideally suited to treat magnetic order and disorder tendencies on an unbiased footing~\cite{reuther-10prb144410,reuther-11prb024402,reuther-11prb014417,reuther-11prb100406,reuther-12prb155127}. Our objectives are two-fold. First, we obtain a detailed understanding of the $J_1$-$J_2$ quantum phase diagram, including all magnetically ordered and disordered regimes as well as the associated phase transitions. In particular, in light of new findings for compounds such as Kapellasite, we allow both $J_1$ and $J_2$ couplings to be ferromagnetic and antiferromagnetic. Aside from ferromagnetic, cuboc, $\sqrt{3}\times \sqrt{3}$ and $\bs{q}=0$ order, we find magnetically disordered regimes located around $(J_1,J_2)\sim (1,0)$, $(J_1,J_2) \sim (0,1)$, and possibly a nonmagnetic phase separating the cuboc from the ferromagnetic domain (Fig.~\ref{fig:magn_order}). Second, we specifically investigate the $J_1$-KHM disordered regime which is supposed to relate to the Herbertsmithite scenario. Due to the large system sizes of up to $317$ sites which are reached by PFFRG, we obtain accurate resolution of the momentum-resolved static magnetic susceptibility. Aside from short-range correlations, we observe a broad spectral distribution which  becomes flat at the magnetic zone boundary for $J_2/J_1 \sim 0.017$ (Fig.~\ref{fig:kagome_nn}). This profile is similar to what is observed in recent neutron measurements~\cite{han-12n406}. Furthermore, we compute the dimer susceptibilities for different pattern candidates. While local pinwheel correlations tend to get enhanced by the RG flow, long-range VBC orders are rejected as seen by a strongly inhomogeneous pattern response (Fig.~\ref{fig:dimer}). This suggests that the initial RG bias for VBC flows away towards a random valence bond (RVB) liquid type state, which is consistent with spin liquid proposals for the KHM.  

{\it Model.} The Hamiltonian of the $J_1$-$J_2$ kagome Heisenberg model (KHM) is given by
\begin{equation}
H_{\text{KHM}}=J_1\sum_{\langle ij\rangle}{\bf S}_i{\bf S}_j+J_2\sum_{\langle\langle ij\rangle\rangle}{\bf S}_i{\bf S}_j, \label{ham}
\end{equation}
where $\langle ij\rangle$ and $\langle\langle ij\rangle\rangle$ denote nearest neighbor  and second neighbor pairs, respectively.
We parametrize the couplings by $J_1=J\cos\alpha$ and $J_2=J\sin\alpha$ which enables us to characterize each point in parameter space by a single angle $\alpha$ (Fig.~\ref{fig:magn_order}), with $0\leq\alpha<2\pi$. 
For $\alpha=90^\circ$, the system decouples into three independent Kagome lattices which implies that the system at this point has the same physical properties as at $\alpha=0$. For its classical counterpart~\cite{domenge-05prb024433}, the $J_1$-$J_2$  KHM exhibits four types of magnetic order: (i) the planar ${\bf q}=0$-N\'eel state with a 3 site unit cell which appears for purely antiferromagnetic interactions at $0^\circ<\alpha<90^\circ$, (ii) the non-planar cuboc state with a $12$ site unit cell located at $90^\circ<\alpha<161.6^\circ$, (iii) a ferromagnetic phase at $161.6^\circ<\alpha<270^\circ$ and (iv) the planar $\sqrt{3}\times\sqrt{3}$-N\'eel state with a 9 site unit cell in the region $270^\circ<\alpha<360^\circ$ (the phases are also found in the quantum phase diagram Fig.~\ref{fig:magn_order}). These types of order correspond to different ordering-peak positions in the second Brillouin zone~\cite{aux}.  
For real-space illustrations of the different orders, we refer to the Refs.~\onlinecite{domenge-05prb024433,messio-11prb184401}. 

{\it Pseudofermion functional renormalization group.} The PFFRG approach~\cite{reuther-10prb144410,reuther-11prb024402,reuther-11prb014417,reuther-11prb100406}, which we employ to obtain the quantum phase diagram displayed in Fig.~\ref{fig:magn_order}, starts by reformulating the spin Hamiltonian in terms of a pseudo fermion representation of the spin-1/2 operators $S^{\mu} = 1/2 \sum_{\alpha\beta} f_{\alpha}^{\dagger} \sigma_{\alpha\beta}^{\mu} f_{\beta}$, ($\alpha,\beta = \uparrow,\downarrow$, $\mu = x,y,z$) with fermionic operators $f_{\uparrow}$ and $f_{\downarrow}$ and Pauli-matrices $\sigma^{\mu}$. This allows to apply Wick's theorem leading to standard Feynman many-body techniques. 
We further introduce an infrared frequency cutoff $\Lambda$ in the fermionic propagator. The FRG ansatz then formulates equations for the evolution of all $m$-particle vertex functions~\cite{metzner-12rmp299} under the flow of $\Lambda$. 
To reduce the infinite hierarchy of coupled equations to a closed set, the PFFRG only includes 2-particle reducible two-loop contributions~\cite{katanin04prb115109}, which still assures a sufficient back-feeding of the self-energy corrections to the two-particle vertex evolution. A crucial advantage of the PFFRG as compared to Abrikosov-type spin-RPA methods is that the summations include vertex corrections between all interaction channels, i.e., the two-particle vertex includes graphs that favor magnetic order and those that favor disorder in such a way that it treats both tendencies on an equal footing. 
A numerical solution of the PFFRG equations requires i) to discretize the frequency dependencies
and ii) to limit the spatial dependence to a finite cluster.
In our calculations, the latter typically includes a correlation area (cluster size) of $317$ lattice sites of the kagome lattice. The onset of spontaneous long-range order is signaled by a sudden breakdown of the smooth RG flow, while the existence of a stable solution indicates the absence of long-range order~\cite{reuther-10prb144410,reuther-11prb024402}. From the effective low-energy two particle vertex, we obtain the spin susceptibility with a high momentum resolution (Fig.~\ref{fig:magn_order} and Fig.~\ref{fig:kagome_nn}) and, in case of a magnetically disordered regime, track possible VBC orders~\cite{reuther-11prb014417} by computing the response functions to different dimer patterns (Fig.~\ref{fig:dimer}).

{\it Quantum phase diagram.}
As shown in Fig.~\ref{fig:magn_order}, we find ferromagnetic, cuboc, $\bs{q}=0$, and $\sqrt{3}\times\sqrt{3}$-N\'eel order in the quantum $J_1$-$J_2$ KHM. 
Fig.~\ref{fig:magn_order} (bottom center) shows the RG flow in the four magnetically ordered phases. Clear instability breakdowns as resolved by strong kinks are seen in the ferromagnetic and in the $\sqrt{3}\times\sqrt{3}$-N\'eel phase. In the ${\bf q}=0$-N\'eel phase and in the cuboc phase, however, the instability features are less pronounced, which is consistent with a small magnetization. We conclude that disorder fluctuations are still important in these regions. Fig.~\ref{fig:magn_order} shows susceptibility profiles of the magnetically ordered phases (the plots are taken for the cutoff value $\Lambda_c$ at the instability breakdown). It exhibits distinct peak structures as expected for the different orders~\cite{aux}. Note that for the ferromagnetic and the $\sqrt{3}\times\sqrt{3}$-N\'eel phase, the PFFRG also accurately resolves the expected subdominant peaks. 

Due to quantum fluctuations, magnetically disordered phases complement the ordered phases in the $J_1$-$J_2$ KHM quantum phase diagram. We find a non-magnetic region 
for $338^\circ \lesssim \alpha \lesssim 35^\circ$ 
(Fig.~\ref{fig:magn_order}). A similar non-magnetic phase is found for $56^\circ \lesssim \alpha \lesssim 124^\circ$.
 In addition, we find indication for a small disordered phase at $\alpha \approx 161$. Enhanced uncertainties near the phase boundary between the cuboc and the ferromagnetic phase, however, make it hard to resolve the range of this phase.  

The comparably small regime of unambiguous cuboc order is an important information in light of recent experimental findings for Kapellasite~\cite{flaak-12prl037208}. Our results confirm that the only way to accomplish such a phase in the quantum $J_1$-$J_2$ KHM is the existence of a ferromagnetic $J_1$ and a considerable antiferromagnetic $J_2$, while even longer-range Heisenberg couplings might additionally be important. In particular, we find the cuboc phase in a regime with enhanced quantum fluctuations, suggesting that comparably small changes of system parameters should induce a significant change in $T_\text{N}$ and the general magnetic susceptibility profile. The PFFRG is ideally suited to track the evolution of the susceptibility peaks in the cuboc domain, which we defer to a later point.

{\it Disordered phase for $\vert \frac{J_2}{J_1} \vert \ll 1 $.} The $J_1$ KHM ($\alpha=0$) has been frequently studied in the literature, and is supposed to be close to the parameter regime where Herbertsmithite is located.  There, we find that the RG flow remains stable in the entire flow regime, such that the RG equations can be smoothly integrated down to $\Lambda=0$. Fig.~\ref{fig:kagome_nn}a depicts the largest susceptibility component (i.e., the one at the K points of the second Brillouin zone). No sign of any instability breakdown is seen during the flow, which is a signature for a non-magnetic ground state. This property along with the indication for a spin gap is also reflected in the spin-spin correlations (inset of Fig.~\ref{fig:kagome_nn}a). Our data show an almost perfect exponential decay of the correlations in real space, from which we determine a small correlation length of only $\xi=0.98$ lattice constants in correspondence with DMRG calculations~\cite{depenbrock-12prl067201}. The full $\bs{k}$-space resolved susceptibility in Fig.~\ref{fig:kagome_nn}b reveals further information about the magnetic properties at $\alpha=0$. The susceptibility is mainly concentrated at the boundary of the second Brillouin zone and is almost constant along the whole edge. Fluctuations at such large $\bs{k}$ are consistent with the small correlation length. The susceptibility along the edge shows only small maxima at the K points, which have also been found in DMRG calculations~\cite{depenbrock-12prl067201}. As a result, subleading $\sqrt{3}\times\sqrt{3}$ fluctuations are slightly preferred as compared to other fluctuations at the Brillouin zone boundary. 
Most of the discovered features such as a broad susceptibility profile are shared by neutron scattering susceptibility measurements in Herbertsmithites, and have been interpreted as signs of spin fractionalization which can emerge in various spin liquid scenarios~\cite{han-12n406}. One major discrepancy, however, is the fact that the measurements do not find any maxima at the K points. By increasing $J_2$, we discover that the previous maxima at $\alpha=0$ vanish and look similar as the experimental finding: Fig.~\ref{fig:kagome_nn}(c) shows the spin susceptibility at $J_2/J_1=0.017$, indicating that the absence of the boundary peaks in Herbertsmithite can be explained by the presence of a small, but finite $J_2>0$ coupling. 

The natural competitor for a spin liquid in such a magnetically disordered phase is VBC order.  For $\alpha=0$, we have considered generalized susceptibilities which measure the propensity of the system to form a specific VBC. As has been shown by dimer expansions~\cite{singh-07prb180407}, the honeycomb VBC (HVBC) (Fig.~\ref{fig:dimer}a) is the most promising dimerization pattern for the $J_1$ KHM. It is a longstanding question whether or not the ground state of this system is a HVBC.
In order to calculate valence bond susceptibilities for such dimer coverings within our RG approach, we add a small perturbation $H_{\text{D}}$ to the model in~\eqref{ham} which increases ($J_1\rightarrow J_1+\delta$) or decreases ($J_1\rightarrow J_1-\delta$) the couplings on the nearest neighbor bonds, according to the dimer pattern. If the system supports the dimer pattern, the strong bonds become stronger and the weak bonds become weaker during the flow~\cite{aux}. 

The RG flow of the HVBC pattern is depicted in Fig.~\ref{fig:dimer}a. The left picture denotes the modified bond strengths at the start of the RG flow. The right picture shows the bond strengths after the flow at $\Lambda=0$. As is clearly visible, the pinwheel bonds around the rotation center of the HVBC structure are strongly and homogeneously enhanced. This has been noticed before, as the pinwheel is an approximate local eigenstate for the KHM~\cite{singh-07prb180407}. For the complete HVBC, however, we find decreasing and less homogeneous response profiles as we consider the bonds near the boundary of the so-defined HVBC unit cell. As a further attempt, we have investigated a modified HVBC pattern where the pinwheel structure stays unchanged, but the arrangement of the outer bonds is modified (Fig.~\ref{fig:dimer}b). The results we obtain show the same degree of inhomogeneous response as for the HVBC. 
Keeping the pinwheel structure, we have also investigated further possible VBC patterns~\cite{aux}.
The unified picture emerging from our studies is that while the pinwheel structures are reasonable guesses for the local correlation profile in the KHM, we do not find any sign of long-range VBC order, as we only find inhomogeneous dimer responses for any VBC pattern considered. More so, the finding of many competing VBC pattern candidates frustrating each other closely resembles the short-range RVB liquid hypothesis, which can seed the emergence of spin liquid physics~\cite{kivelson-87prb8865}.

To summarize, we have obtained the quantum phase diagram of the $J_1$-$J_2$ KHM and investigated its magnetic and non-magnetic phases. In light of recent experiments on Kapellasite, it will be interesting to further investigate the interplay of order and disorder in the cuboc order regime. Furthermore, we have provided model evidence for explaining the features of the susceptibility measurements in Herbertsmithite, and specifically link the absence of boundary susceptibility enhancements to the existence of small but finite $J_2$ coupling. We have also found indications for RVB liquid physics in the $J_1$ KHM, rendering the related Herbertsmithite a prime candidate for observing a spin liquid phase in nature.

\begin{acknowledgements}
RT thanks A.~C.~Potter, D.~Mross, C.~Lhuillier, A.~L\"auchli, P.~A.~Lee, F.~Mila, and all participants of the KITP workshop "Frustrated Magnetism and Quantum Spin Liquids: From Theory and Models to Experiments''. CP and RT are supported by SPP 1458. JR acknowledges support from the Deutsche Akademie der Naturforscher Leopoldina through grant LPDS 2011-14.
\end{acknowledgements}


\begin{thebibliography}{10}

\bibitem{harris-92prb2899}
A.~B. Harris, C. Kallin, and A.~J. Berlinsky, Phys. Rev. B {\bf 45},  2899
  (1992).

\bibitem{chalker-92prl855}
J.~T. Chalker, P.~C.~W. Holdsworth, and E.~F. Shender, Phys. Rev. Lett. {\bf
  68},  855  (1992).

\bibitem{anderson87s1196}
P.~W. Anderson, Science {\bf 235},  1196  (1987).

\bibitem{marston-91jap5962}
J.~B. Marston and C. Zeng, J. Appl. Phys. {\bf 69},  5962  (1991).

\bibitem{wen91prb2664}
X.~G. Wen, Phys. Rev. B {\bf 44},  2664  (1991).

\bibitem{sachdev92prb12377}
S. Sachdev, Phys. Rev. B {\bf 45},  12377  (1992).

\bibitem{hastings00prb014413}
M.~B. Hastings, Phys. Rev. B {\bf 63},  014413  (2000).

\bibitem{ran-07prl117205}
Y. Ran, M. Hermele, P.~A. Lee, and X.-G. Wen, Phys. Rev. Lett. {\bf 98},
  117205  (2007).

\bibitem{waldtmann98epj501}
C. Waldtmann, H.~U. Everts, B. Bernu, C. Lhuillier, P. Sindzingre, P.
  Lecheminant, and L. Pierre, Eur. Phys. J. B {\bf 2},  501  (1998).

\bibitem{jiang-08prl117203}
H.~C. Jiang, Z.~Y. Weng, and D.~N. Sheng, Phys. Rev. Lett. {\bf 101},  117203
  (2008).

\bibitem{yan-11s1173}
S. Yan, D.~A. Huse, and S.~R. White, Science {\bf 332},  1173  (2011).

\bibitem{depenbrock-12prl067201}
S. Depenbrock, I.~P. McCulloch, and U. Schollw\"ock, Phys. Rev. Lett. {\bf
  109},  067201  (2012).

\bibitem{iqbal-cm1209}
Y. Iqbal, F. Becca, S. Sorella, and D. Poilblanc, arXiv:1209.1858.

\bibitem{helton-07prl107204}
J.~S. Helton, K. Matan, M.~P. Shores, E.~A. Nytko, B.~M. Bartlett, Y. Yoshida,
  Y. Takano, A. Suslov, Y. Qiu, J.-H. Chung, D.~G. Nocera, and Y.~S. Lee, Phys.
  Rev. Lett. {\bf 98},  107204  (2007).

\bibitem{mendels-07prl077204}
P. Mendels, F. Bert, M.~A. de~Vries, A. Olariu, A. Harrison, F. Duc, J.~C.
  Trombe, J.~S. Lord, A. Amato, and C. Baines, Phys. Rev. Lett. {\bf 98},
  077204  (2007).

\bibitem{elhajal-02prb014422}
M. Elhajal, B. Canals, and C. Lacroix, Phys. Rev. B {\bf 66},  014422  (2002).

\bibitem{devries-08prl157205}
M.~A. de~Vries, K.~V. Kamenev, W.~A. Kockelmann, J. Sanchez-Benitez, and A.
  Harrison, Phys. Rev. Lett. {\bf 100},  157205  (2008).

\bibitem{helton-10prl147201}
J.~S. Helton, K. Matan, M.~P. Shores, E.~A. Nytko, B.~M. Bartlett, Y. Qiu,
  D.~G. Nocera, and Y.~S. Lee, Phys. Rev. Lett. {\bf 104},  147201  (2010).

\bibitem{han-12prl157202}
T. Han, S. Chu, and Y.~S. Lee, Phys. Rev. Lett. {\bf 108},  157202  (2012).

\bibitem{wulferding-10prb144412}
D. Wulferding, P. Lemmens, P. Scheib, J. R\"oder, P. Mendels, S. Chu, T. Han,
  and Y.~S. Lee, Phys. Rev. B {\bf 82},  144412  (2010).

\bibitem{han-12n406}
T.-H. Han, J.~S. Helton, S. Chu, D.~G. Nocera, J.~A. Rodriguez-Rivera, C.
  Broholm, and Y.~S. Lee, Nature {\bf 492},  406  (2012).

\bibitem{coldea-03prb134424}
R. Coldea, D.~A. Tennant, and Z. Tylczynski, Phys. Rev.~B {\bf 68},  134424
  (2003).

\bibitem{flaak-12prl037208}
B. F\aa{}k, E. Kermarrec, L. Messio, B. Bernu, C. Lhuillier, F. Bert, P.
  Mendels, B. Koteswararao, F. Bouquet, J. Ollivier, A.~D. Hillier, A. Amato,
  R.~H. Colman, and A.~S. Wills, Phys. Rev. Lett. {\bf 109},  037208  (2012).

\bibitem{janson-08prl106403}
O. Janson, J. Richter, and H. Rosner, Phys. Rev. Lett. {\bf 101},  106403
  (2008).

\bibitem{reuther-10prb144410}
J. Reuther and P. W\"olfle, Phys. Rev. B {\bf 81},  144410  (2010).

\bibitem{reuther-11prb024402}
J. Reuther and R. Thomale, Phys. Rev. B {\bf 83},  024402  (2011).

\bibitem{reuther-11prb014417}
J. Reuther, D.~A. Abanin, and R. Thomale, Phys. Rev. B {\bf 84},  014417
  (2011).

\bibitem{reuther-11prb100406}
J. Reuther, R. Thomale, and S. Trebst, Phys. Rev. B {\bf 84},  100406  (2011).

\bibitem{reuther-12prb155127}
J. Reuther, R. Thomale, and S. Rachel, Phys. Rev. B {\bf 86},  155127  (2012).

\bibitem{domenge-05prb024433}
J.-C. Domenge, P. Sindzingre, C. Lhuillier, and L. Pierre, Phys. Rev. B {\bf
  72},  024433  (2005).

\bibitem{aux}
For further details see the supplementary material.

\bibitem{messio-11prb184401}
L. Messio, C. Lhuillier, and G. Misguich, Phys. Rev. B {\bf 83},  184401
  (2011).

\bibitem{metzner-12rmp299}
W. Metzner, M. Salmhofer, C. Honerkamp, V. Meden, and K. Sch\"onhammer, Rev.
  Mod. Phys. {\bf 84},  299  (2012).

\bibitem{katanin04prb115109}
A.~A. Katanin, Phys. Rev. B {\bf 70},  115109  (2004).

\bibitem{singh-07prb180407}
R.~R.~P. Singh and D.~A. Huse, Phys. Rev. B {\bf 76},  180407  (2007).

\bibitem{kivelson-87prb8865}
S.~A. Kivelson, D.~S. Rokhsar, and J.~P. Sethna, Phys. Rev. B {\bf 35},  8865
  (1987).

\end{thebibliography}

\newpage
\begin{center}
{\large \bf Supplementary material}
\end{center}

\section{Classical phase diagram}
\begin{figure}[t]
$\begin{matrix} \begin{matrix}\includegraphics[scale=0.78]{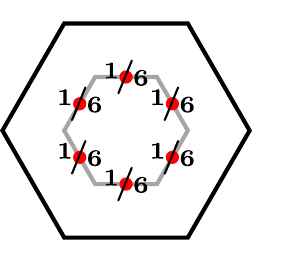} \\ \includegraphics[scale=0.78]{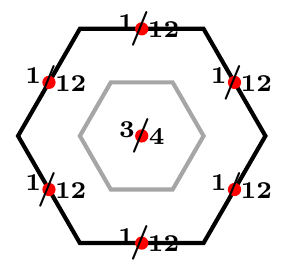}\hspace*{0.2cm} \end{matrix} \!\! \begin{matrix} \includegraphics[scale=0.78]{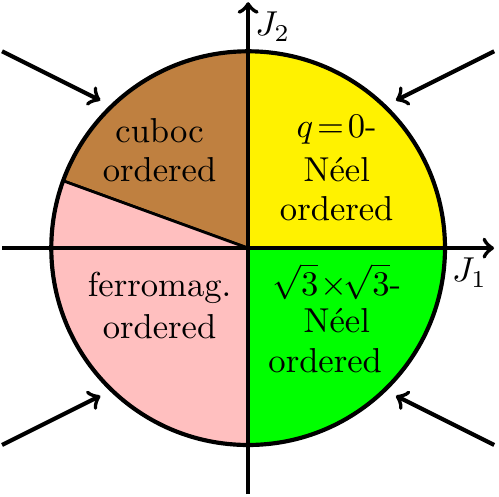} \end{matrix} \!\! \begin{matrix}\includegraphics[scale=0.78]{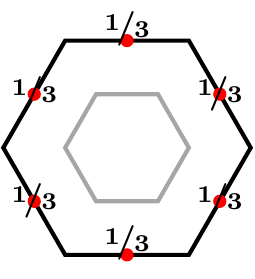} \\ \includegraphics[scale=0.78]{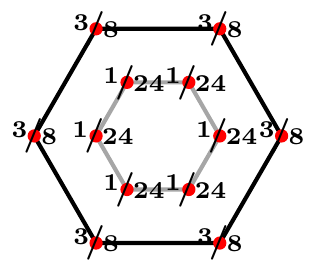} \end{matrix} \end{matrix}$
\caption{The phase diagram of the classical $J_1$-$J_2$ KHM with ${\bf q}=0$ order, cuboc order, ferromagnetic order, and $\sqrt{3}\times\sqrt{3}$ order. The left and right sides of the figure show the $\bs{k}$-space positions of the classical ordering peaks corresponding to these types of order (red dots). The black (gray) hexagons represent the boundaries of the second (first) Brillouin zone. Numbers indicate the relative peak heights of the different ordering peaks, which always add up to one within the second Brillouin zone.}
\label{fig:classical}
\end{figure}

In the following, we discuss the classical $J_1$-$J_2$ KHM as obtained in the large-spin limit in more detail. Fig.~\ref{fig:classical} shows the corresponding phase diagram comprising ${\bf q}=0$ order, cuboc order, ferromagnetic order, and $\sqrt{3}\times\sqrt{3}$ order. All phase boundaries coincide either with the $J_1$ or the $J_2$ axis, except for the transition between the cuboc and the ferromagnetic phase, which lies at $J_2=-J_1/3$. The peak structures of the Fourier transforms of these types of order are also shown in Fig.~\ref{fig:classical}. We emphasize that Fourier transformed quantities on the Kagome lattice have the periodicity of the second (extended) Brillouin zone, indicated by black hexagons in Fig.~\ref{fig:classical}. Unlike the cuboc and ${\bf q}=0$ order, the ferromagnetic and $\sqrt{3}\times\sqrt{3}$ order cannot be decomposed into harmonics with equal wave vectors $|{\bs k}|$. These types of order require two inequivalent harmonics, where the dominant (subdominant) contribution resides at the boundary of the second Brillouin zone (inside the second Brillouin zone).

\section{Dimer correlations}
\begin{figure}[t]
\centering
\hspace*{-7.15cm}$\begin{matrix} \begin{matrix}\includegraphics{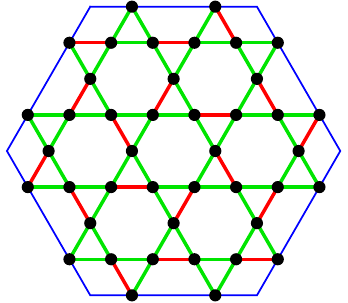}\end{matrix} \to \begin{matrix}\includegraphics{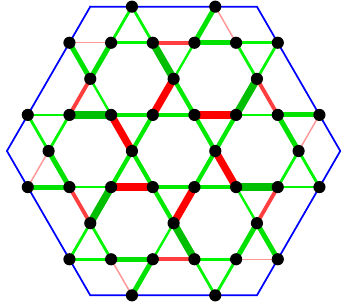}\end{matrix}\!\!\!\!\!\!\!\!\!\!\!\!\!\!\!\!\!\!\!\!\!\!\!\!\!\!\!\!\!\!\!\!\!\!\!\!\!\!\!\!\!\!\!\!\!\!\!\!\!\!\!\!\!\!\!\!\!\!\!\!\!\!\!\!\!\!\!\!\!\!\!\!\!\!\!\!\!\!\!\!\!\!\!\!\!\!\!\!\!\!\!\!\!\!\!\!\!\!\!\!\!\!\!\!\!\!\!\!\!\!\!\!\!\!\!\!\!\!\!\!\!\!\!\!\!\!\!\!\!\!\!\!\!\!\!\!\!\!\!\!\!\! \begin{matrix}\text{(a)}\\\vphantom{m}\\\vphantom{m}\\\vphantom{m}\\\vphantom{m}\\\vphantom{m}\\\vphantom{m}\end{matrix}\end{matrix}$\\[0.4cm]
$\begin{matrix} \begin{matrix}\includegraphics{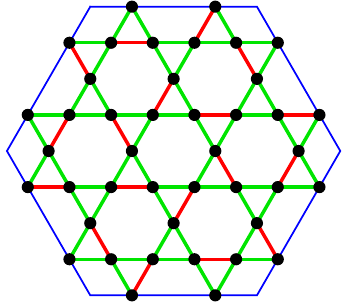}\end{matrix} \to \begin{matrix}\includegraphics{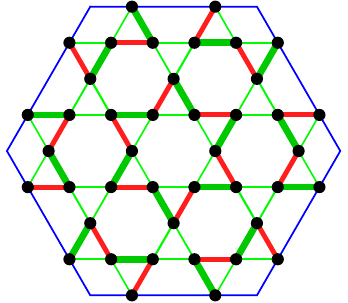}\end{matrix}\!\!\!\!\!\!\!\!\!\!\!\!\!\!\!\!\!\!\!\!\!\!\!\!\!\!\!\!\!\!\!\!\!\!\!\!\!\!\!\!\!\!\!\!\!\!\!\!\!\!\!\!\!\!\!\!\!\!\!\!\!\!\!\!\!\!\!\!\!\!\!\!\!\!\!\!\!\!\!\!\!\!\!\!\!\!\!\!\!\!\!\!\!\!\!\!\!\!\!\!\!\!\!\!\!\!\!\!\!\!\!\!\!\!\!\!\!\!\!\!\!\!\!\!\!\!\!\!\!\!\!\!\!\!\!\!\!\!\!\!\!\! \begin{matrix}\text{(b)}\\\vphantom{m}\\\vphantom{m}\\\vphantom{m}\\\vphantom{m}\\\vphantom{m}\\\vphantom{m}\end{matrix}\qquad\qquad\qquad\qquad\qquad\qquad\qquad\qquad\qquad\qquad\qquad \end{matrix}$\\[0.4cm]
$\begin{matrix} \begin{matrix}\includegraphics{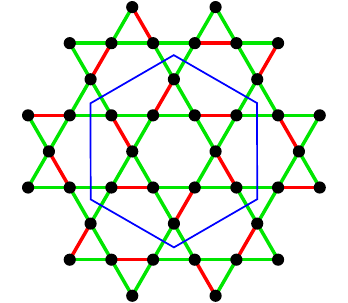}\end{matrix} \to \begin{matrix}\includegraphics{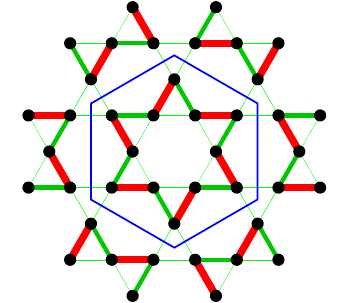}\end{matrix}\!\!\!\!\!\!\!\!\!\!\!\!\!\!\!\!\!\!\!\!\!\!\!\!\!\!\!\!\!\!\!\!\!\!\!\!\!\!\!\!\!\!\!\!\!\!\!\!\!\!\!\!\!\!\!\!\!\!\!\!\!\!\!\!\!\!\!\!\!\!\!\!\!\!\!\!\!\!\!\!\!\!\!\!\!\!\!\!\!\!\!\!\!\!\!\!\!\!\!\!\!\!\!\!\!\!\!\!\!\!\!\!\!\!\!\!\!\!\!\!\!\!\!\!\!\!\!\!\!\!\!\!\!\!\!\!\!\!\!\!\!\! \begin{matrix}\text{(c)}\\\vphantom{m}\\\vphantom{m}\\\vphantom{m}\\\vphantom{m}\\\vphantom{m}\\\vphantom{m}\end{matrix}\qquad\qquad\qquad\qquad\qquad\qquad\qquad\qquad\qquad\qquad\qquad \end{matrix}$
\caption{Further valence-bond configurations that we have studied. Again, red lines correspond to strengthened and green lines to weakened bonds. Blue lines mark the unit cell boundaries (note the smaller unit cell in (c)). The pattern in (a) is similar to the valence bond configurations in Fig.~\ref{fig:dimer}, i.e., it exhibits "perfect hexagons" at the unit cell corners. (b) represents a pattern of clockwise and counterclockwise rotating pinwheels. A pattern of pinwheels with the same sense of rotation is shown in (c).}
\label{fig:dimer2}
\end{figure}

In order to investigate the properties of the magnetically disordered regimes of the quantum $J_1$-$J_2$ KHM we have probed the system with respect to its propensity to form various dimerized VBCs. By definition, a VBC is a periodic arrangement of dimer bonds on pairs of nearest neighbor sites $(i,j)$, such that each site belongs to exactly {\it one} dimer. In the following, for a given configuration, $S$ denotes the set of all such pairs of sites while $W$ is the set of all other nearest neighbor bonds. Within our FRG framework, a conceptually simple way to calculate generalized dimer susceptibilities $\chi^{\text{dimer}}$, measuring the tendency of the system to support a specific VBC, is to add a small perturbation $H_{\text{D}}$ to the Hamiltonian,
\begin{equation}
H_{\text{D}}=\delta\sum_{(i,j)\in S}{\mathbf S}_i{\mathbf S}_j-\delta\sum_{(i,j)\in W}{\mathbf S}_i{\mathbf S}_j\,,
\end{equation}
which strengthens the couplings $J_{ij}$ on all dimer bonds in $S$ [$J_{ij}\rightarrow J_{ij}+\delta$ if $(i,j)\in S$] and weakens all other nearest neighbor couplings [$J_{ij}\rightarrow J_{ij}-\delta$ if $(i,j)\in W$]. Such a modification affects the initial conditions of the RG flow at large cutoff scales $\Lambda$. As $\Lambda$ is lowered, we keep track of the evolution of all static nearest neighbor spin-spin correlations $\chi_{ij}=\langle\langle {\mathbf S}_i{\mathbf S}_j \rangle\rangle(\omega=0)$. We then define the dimer susceptibility for a given pair of adjacent sites $(i,j)$ by $\chi^{\text{dimer}}_{ij}=\chi_{ij}-\chi_m$ where $\chi_m$ is a properly chosen mean value over all nearest neighbor bonds (see below). If the absolute value $|\chi^{\text{dimer}}_{ij}|$ is small, the system tends to equalize (i.e. reject) the perturbation on that link, while a large value indicates that the system supports the dimerization. In Figs.~\ref{fig:dimer} and \ref{fig:dimer2}, $|\chi^{\text{dimer}}_{ij}|$ is encoded in the thickness of the lines connecting nearest neighbor bonds. Furthermore, the color (green, red) indicates the the sign of $\chi^{\text{dimer}}_{ij}$ which distinguishes between weakened and strengthened bonds.

For the HVBC introduced in the paper, as well as for most of the other patterns that we have tested, the unit cell consists of 6 inequivalent strengthened nearest neighbor bonds with $(i,j)\in S$ and 18 inequivalent weakened nearest neighbor bonds with $(i,j)\in W$. It is convenient to define the mean value of the spin-spin correlations by
\begin{equation}
\chi_m=\left(\sum_{(i,j)\in S} 3\chi_{ij}+\sum_{(i,j)\in W} \chi_{ij}\right)/36\,,\label{mean_value}
\end{equation}
where the sums only run over inequivalent links in a unit cell. In Eq.~(\ref{mean_value}) the strengthened bonds contribute with an additional factor of $3$. This ensures that in total strengthened and weakened bonds are equally weighted such that at the beginning of the RG flow (i.e., at large $\Lambda$) $|\chi^{\text{dimer}}_{ij}|$ is constant on all bonds (left-hand pictures in Figs.~\ref{fig:dimer} and \ref{fig:dimer2}). In general, not only the $\Lambda$-evolution of the magnitude of the dimer susceptibilities $|\chi^{\text{dimer}}_{ij}|$ is relevant but also their spatial homogeneity: The formation of a VBC is indicated by a large {\it and} homogeneous dimer response. On the other hand, an inhomogeneous response shows that certain bonds might have a bias towards dimerization while the dimer pattern as a whole is rejected and long-range order does not develop.

In the main text, we have discussed the HVBC as well as a slightly modified pattern (Fig.~\ref{fig:dimer}). While the central pinwheel structure gains considerable weight during the RG flow, the overall response is rather inhomogeneous. Keeping the 36 site unit cell and the pinwheel, there exists another related pattern, which we have also tested, see Fig.~\ref{fig:dimer2}a. The susceptibility distribution shows the same degree of inhomogeneity as in Fig.~\ref{fig:dimer}. Note that the corners of the unit cells in Figs.~\ref{fig:dimer} and \ref{fig:dimer2}a exhibit a local dimer arrangement which has been referred to as "perfect hexagons" (i.e. hexagons with three dimer bonds along its edges). Obviously these features represent the main source of small and inhomogeneous responses in Figs.~\ref{fig:dimer} and \ref{fig:dimer2}a.

Given these results it is interesting to study valence bond configurations which only consist of pinwheels and {\it no} "perfect hexagons". Fig.~\ref{fig:dimer2}b shows such a pattern where each strengthened dimer bond belongs to a pinwheel (this is best seen by putting together various unit cells). Together with Figs.~\ref{fig:dimer} and \ref{fig:dimer2}a, these four patterns are the only possible valence bond configurations with a $36$ site unit cell, a pinwheel structure and (at least) $120^\circ$ rotation symmetry. Note that as in Fig.~\ref{fig:dimer2}b, pinwheels have a handedness (clockwise and counterclockwise rotating pinwheels occur in the ratio 1:3). Again, while the pinwheels become the salient features during the RG flow, other bonds exhibit small responses, ruling out long-range order of this type.

Finally, we have tested a dimer pattern which only consists of pinwheels with the same handedness, (Fig.~\ref{fig:dimer2}c). This gives a unit cell of only $12$ sites. The response is very similar to Fig.~\ref{fig:dimer2}b, which in total confirms the finding of a low energy scenario involving many competing valence bond configurations.

\end{document}